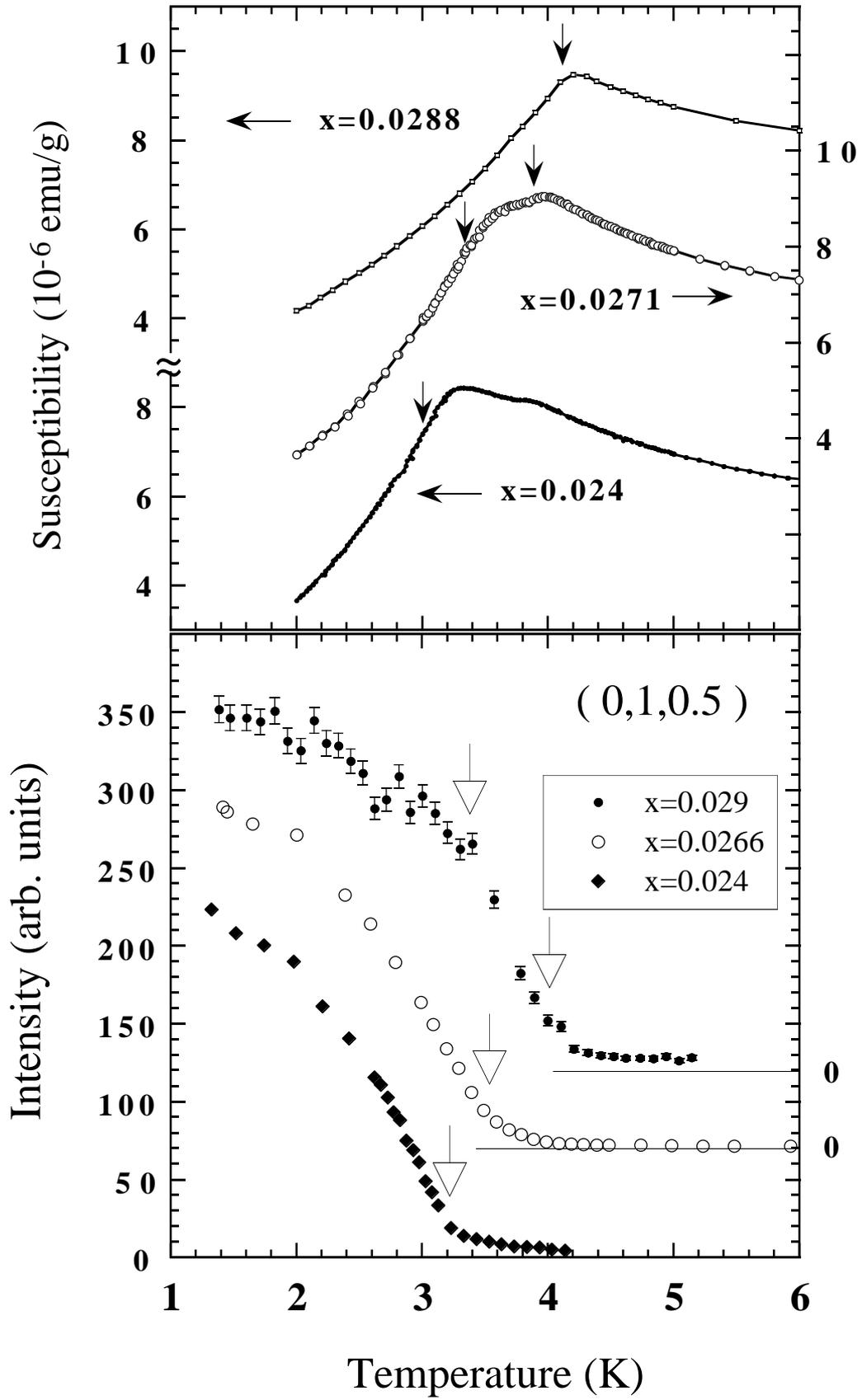

Fig. 1 Nishi et al.

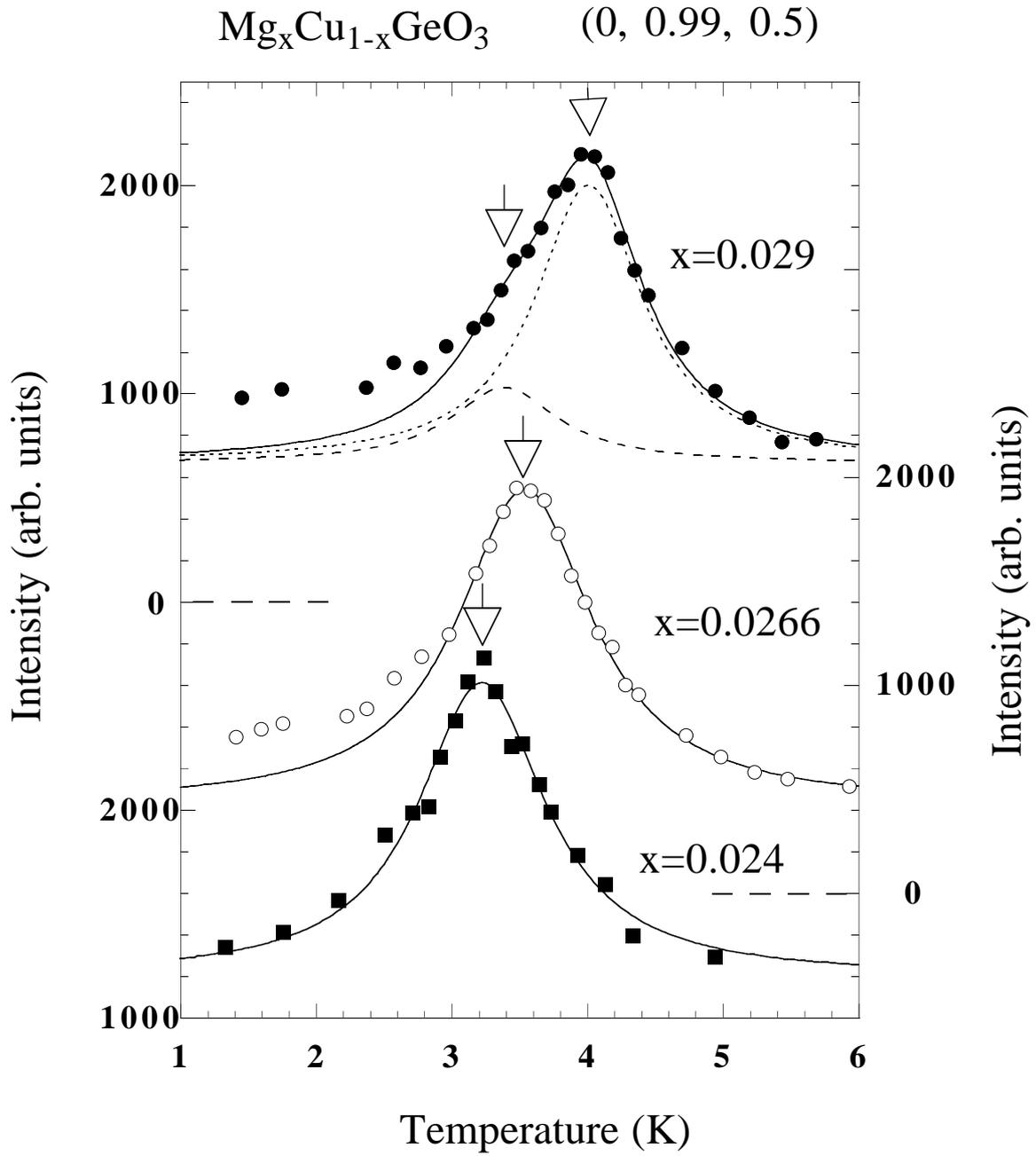

Fig. 2 Nishi et al.

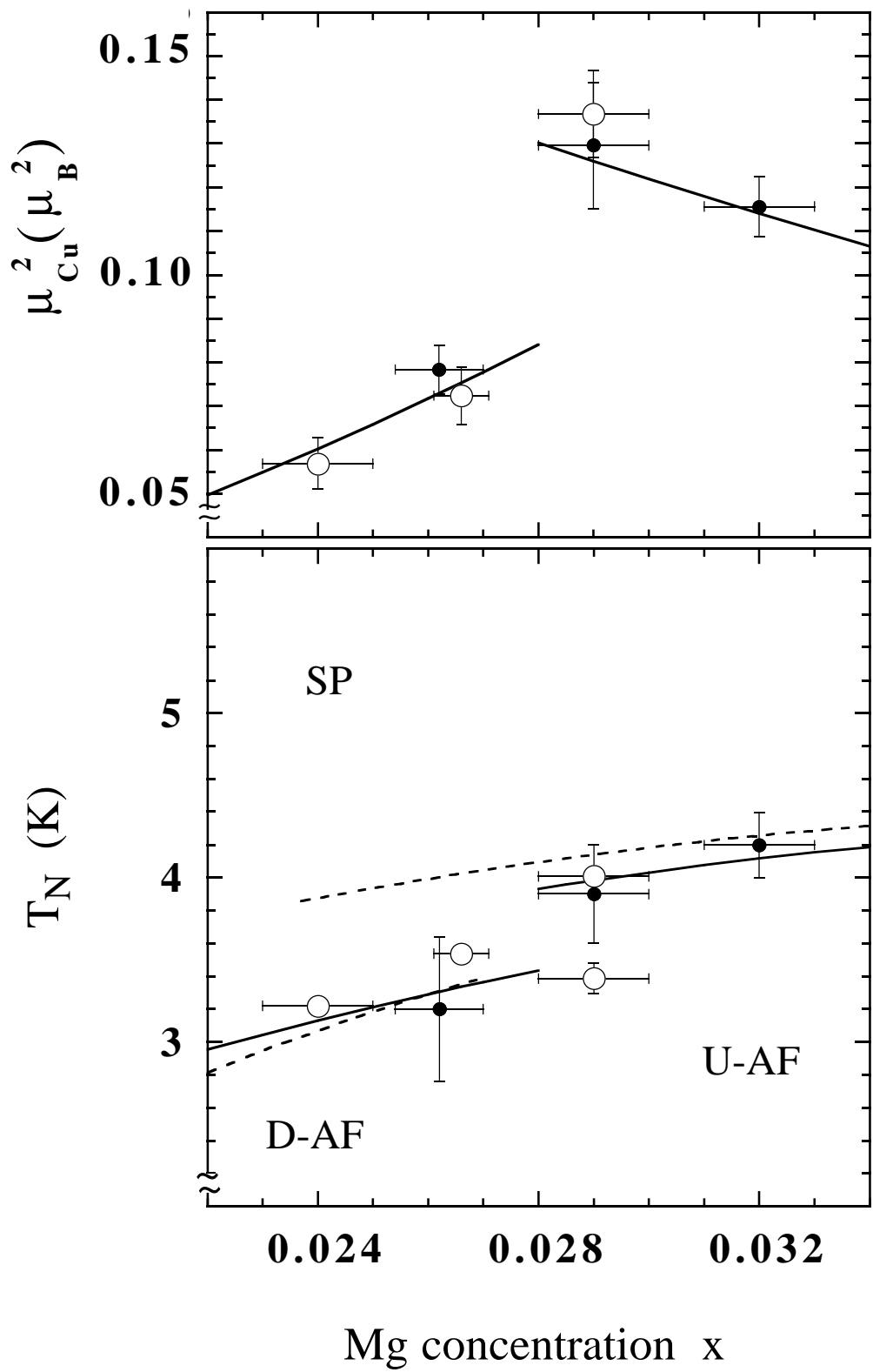

Fig. 3  Nishi et al.

# Critical Neutron Scattering Study of the Compositional Phase Transition in Mg-Doped CuGeO$_3$

Masakazu NISHI*, Hironori NAKAO**, Yasuhiko FUJII, Takatsugu MASUDA[1], Kunimitsu UCHINOKURA[1] and Gen SHIRANE[2]

*Neutron Scattering Laboratory, Institute for Solid State Physics, The University of Tokyo, 106-1 Shirakata, Tokai, Ibaraki, 319-1106*
[1] *Department of Advanced Materials Science, The University of Tokyo, Hongo, Bunkyo-ku, Tokyo, 113-8656*
[2] *Department of Physics, Brookhaven National Laboratory, Upton, NY, 11973-5000, USA*

Cu$_{1-x}$Mg$_x$GeO$_3$ undergoes a first-order phase transition at a critical concentration $x_c$ between an antiferromagnetic (AF) state on dimerized lattice (D-AF) and an AF Néel state on undistorted uniform lattice (U-AF). Previous magnetic susceptibility measurements showed $x_c = 0.023$ while a recent neutron scattering study reported $x_c = 0.027 \pm 0.001$. The present critical scattering due to antiferromagnetic fluctuations near the superlattice reflection (0,1,1/2) unambiguously determines $x_c = 0.028 \pm 0.001$ at $T_N = 3.4 \sim 4$ K. Also at $T = 1.3$ K, the phase boundary was determined as $x_c = 0.028 \pm 0.001$ by observation of a jump of an effective magnetic moment across $x_c$.

KEYWORDS: CuGeO$_3$, spin-Peierls, alternating chains, antiferromagnetism, neutron scattering

The first inorganic spin-Peierls (SP) compound CuGeO$_3$ [1] with $T_{SP}$ = 14 K has offered a great opportunity to study doping effects on its SP state[2]. Regnault *et al.*[3] reported a striking fact that the SP state coexists with antiferromagnetic (AF) state in CuGe$_{1-x}$Si$_x$O$_3$ below $T_N$ ($<T_{SP}$). Fukuyama *et al.*[4] proposed a theoretical model to understand such a coexistence by a unique phase in which both magnetic moments and lattice dimerization are spatially modulated. This AF state stabilized on the dimerized lattice is called D-AF. Recently, Masuda *et al.*[5] reported a first-order phase transition from such a D-AF to a regular AF state on a uniform lattice (called U-AF) at a critical composition $x_c$ = 0.023 in Cu$_{1-x}$Mg$_x$GeO$_3$ by magnetic susceptibility measurements with very carefully controlled Mg concentration of samples. This experimental fact was theoretically understood by Saito[6] who studied impurity effects on SP systems and suggested an occurrence of a first-order phase transition at a critical concentration of impurity at $T$ = 0 K. Recently Nakao *et al.*[7] performed precise measurements of temperature dependence of both antiferromagnetic and SP superlattice peak intensities on Cu$_{1-x}$Mg$_x$GeO$_3$ by neutron scattering to determine its detailed temperature-concentration ($T$-$x$) phase diagram. They clearly observed an abrupt change in both average magnetic moment and atomic displacement due to lattice dimerization at a critical concentration $x_c$ = 0.027 ± 0.001. More recently Masuda *et al.*[8] reported a newer $T$-$x$ phase diagram of the same material by their reanalysis of magnetic susceptibility data. Their susceptibility curve as a function of temperature showed two peaks corresponding to two AF transition temperatures, *i.e.* paramagnetic to D-AF and to U-AF, in the Mg concentration range between 0.0237 and 0.0271.

In this paper, we report a critical neutron scattering study of three kinds of Mg concentration samples near $x_c$, where the critical fluctuations corresponding to two AF transitions are expected to be observed. Critical scattering was clearly was clearly observed in all samples; however, only the sample with $x = 0.029$ indicates such two AF transition temperatures.

Neutron scattering experiments were carried out with the following conditions of an ISSP-owned triple-axis spectrometer (ISSP-GPTAS) installed at JRR-3M of JAERI in Tokai, Japan. Both incident and final neutron energies were fixed at 13.64 meV with (002) reflection of pyrolytic graphite (PG) monochromator and analyzer. The energy resolution was 0.6 meV in FWHM under the horizontal collimation of 40'-20'-20'-40'. PG filters were used to eliminate the higher-order contaminations. A single crystal of Mg-doped $CuGeO_3$ was mounted with the $(0, k, l)$ scattering plane in an ILL-type orange cryostat capable of reaching 1.3 K by pumping liquid helium. A typical mosaicness of each sample measured by neutron scattering was about 0.2 degrees FWHM. The samples used in the present neutron scattering experiment were grown by a floating-zone method and the concentration of Mg was analyzed by an inductively coupled plasma atomic emission spectroscopy (ICP-AES) method in the concentration accuracy of $\pm 0.001$[8]. Thus determined Mg concentrations of these samples were 0.024, 0.0266 and 0.029.

The raw data of magnetic susceptibilities in the applied field parallel to the $c$-axis[8] are shown in Fig. 1 (a). Two data curves show a double-peak structure while peak positions of the previous heat capacity data[9] referred in Table I are shown by vertical arrows in Fig. 1(a). The temperature dependence of neutron scattering intensity of the AF Bragg peak (0, 1, 0.5) is shown in Fig. 1(b) for three kinds of samples to confirm each AF transition temperature ($T_N$). The arrows represent each transition temperature determined from critical scattering shown in Fig. 2. The data of $x = 0.029$ is referred from Fig. 3 of Nakao *et al.*[7].

Magnetic critical scattering was measured by a triple-axis configuration to significantly reduce background. The measurements were performed at $Q$ = (0, 0.99, 0.5) and (0, 0.98, 0.5) for both $x$ = 0.024 and $x$ = 0.0266, and $Q$ = (0, 0.99, 0.5) for $x$ = 0.029. These results are shown in Fig. 2 for $Q$ = (0, 0.99, 0.5). The diverging behavior of the critical scattering was clearly observed at magnetic transition temperature. The peak position of the critical scattering at $x$=0.024 was determined as 3.22 ± 0.02 K and 3.24 ± 0.04 K for $Q$ = (0, 0.99, 0.5) and (0, 0.98, 0.5), respectively, and in the case of $x$=0.026 it was also determined as 3.54 ± 0.02 K and 3.50 ± 0.04 K for each $Q$. Within experimental error, the transition temperature defined as divergence of critical fluctuation at $Q$ = (0, 1, 0.5) can be determined as the same as $Q$ = (0,0.99,0.5). Each transition temperature is determined as $T_N$ =3.22 ± 0.02 K for $x$ = 0.024 and 3.54 ± 0.05 K for $x$ = 0.0266 much more unambiguously than a conventional method to measure a superlattice intensity accompanied with short-range diffuse scattering at $T \geq T_c$. These $T_N$'s correspond to the lower transition temperature, *i.e.* paramagnetic to D-AF. Two AF transition temperatures expected from the magnetic susceptibility[8] and heat capacity[9] were not observable by neutron scattering on the $x$ = 0.0266 sample. On the $x$ = 0.029 sample, on the other hand, a small shoulder peak was observed near the lower $T_N$ while a dominant peak corresponds to the higher AF transition temperature. This curve can be analyzed by two components with two different Néel temperatures of $T_N$ = 4.01 ± 0.02 and 3.39 ± 0.09 as drawn in the figure. These experimental results are consistent with the *T-x* phase diagram of $Cu_{1-x}Mg_xGeO_3$ previously obtained by Nakao *et al.* [7] as displayed in Fig. 3 (bottom), where the present data are plotted with open circles and the previous data with closed one. Mg concentration $x$ of the samples used for the experiment by Nakao *et al.* [7] was analyzed by ICP-AES method, and the result showed the change from $x$ = 0.026 and 0.028 to 0.0262 and 0.029, respectively. Therefore, the compositional phase boundary at $T_N$ = 3.4 ~ 4 K can be determined as $x_c$ = 0.028 ± 0.001. An effective magnetic moment of Cu atoms obtained at $T$ = 1.5 K was obtained from the intensity ratio of the AF reflection (0, 1, 0.5) to the fundamental reflection (0, 2, 0) which was previously confirmed to be extinction free[7]. It is shown at top of Fig. 3 together with the data of Nakao *et al*. One can see the maximum value of the moment just at the phase boundary. The samples used for magnetic susceptibility and heat capacity were obtained from the same batches as those studied in the present neutron scattering experiment. Antiferromagnetic Néel temperatures are tabulated in Table I obtained by three kinds of experimental methods. Different data may be obtained from different experimental methods.

The results of our critical neutron scattering experiments lead us to the conclusion that the compositional phase boundary is determined unambiguously $x_c$ = 0.028 ± 0.001 in confirming the two AF transition temperatures which was not observed by the temperature dependence of the AF Bragg intensity.

We acknowledge M. Saito and H. Fukuyama for fruitful discussions. We thank S. Watanabe for technical support on our experiments. This study was supported in part by the US-Japan Cooperative Research Program on Neutron Scattering between USDOE and MONBUSHO. Work at Brookhaven National Laboratory was carried out under contract No. DE-AC02-98CH10886, Division of Material Science, U. S. Department of Energy.

**Figure captions**

Fig. 1. (a) The magnetic susceptibility data of Mg doped $CuGeO_3$ below 6 K. The data of different $x$ are shifted vertically.  Four vertical arrows show the peak positions of the heat capacity data obtained by Masuda *et al.*[9].
(b) The temperature dependence of the neutron scattering intensities of the (0, 1, 0.5) Bragg peak for three kinds of samples.  White arrows positions are the same as these of critical scattering in Fig. 2.  The data of $x = 0.029$ is referred from Fig. 3 of Nakao *et al.*[7].

Fig. 2.  Critical neutron scattering observed at (0, 0.99, 0.5) near the AF Néel points.   Intensity differences from solid line at low temperature side of $x = 0.0266$ and 0.029 are contributed by the foot of Bragg peak (0, 1, 0.5).  White arrows show the divergence positions of critical scattering intensity.

Fig. 3.  (a) The Mg concentration dependence of the square of magnetic moment of Cu atom $(\mu_{Cu})^2$ at 1.3 K.  Closed circles are referred from Nakao *et al.*[7] and open circles are present data.   Solid curves are guides for eye.
(b) The *T*-*x* magnetic phase diagram based on Mg concentration $x$.  At $x < x_c$ SP state coexists with AF Néel state below the AF ordering temperature $T_N$.  At $x > x_c$ SP state disappears at $T = 0$ K.  The phase boundary between two kinds of Néel states exists at $x_c = 0.028$.  Broken curves are referred from Masuda *et al.* for the magnetic susceptibility data[8].

**TABLE I.** Antiferromagnetic transition temperature (K) of three kinds of samples obtained from the previous magnetic susceptibility ($\chi$) [8] and heat capacity ($C_p$) [9], and the present neutron scattering (NS) data. The small difference of Mg concentrations between upper column and below one are due to different part of the same batches.

| $x$ | 0.024 | | 0.0271 | | 0.0288 | |
|---|---|---|---|---|---|---|
| | $T_{N1}$ | $T_{N2}$ | $T_{N1}$ | $T_{N2}$ | $T_{N1}$ | $T_{N2}$ |
| $\chi$ | 3.25 | 3.87 | 3.57 | 4.02 | - | 4.19 |
| $C_p$ | 3.00 | - | 3.35 | 3.90 | - | 4.12 |

| $x$ | 0.024 | | 0.0266 | | 0.029 | |
|---|---|---|---|---|---|---|
| | $T_{N1}$ | $T_{N2}$ | $T_{N1}$ | $T_{N2}$ | $T_{N1}$ | $T_{N2}$ |
| NS | 3.22 | - | 3.54 | - | 3.39 | 4.01 |

(- not observable)